\documentclass[twocolumn, tighten, numberedappendix, twocolappendix]{aastex631}

\pdfoutput=1 

\usepackage{amsmath,amstext}
\usepackage[T1]{fontenc}
\usepackage{apjfonts} 
\usepackage{graphicx}
\usepackage{xcolor}
\usepackage{color}
\usepackage{appendix}
\usepackage{microtype}
\usepackage{comment}
\usepackage{enumitem}

\newcommand{\WMAP}{WMAP}
\newcommand{\wmap}{WMAP}
\newcommand{\Planck}{{Planck}}
\newcommand{\planck}{{Planck}}

\renewcommand{\ell}{\ensuremath{l}}
\newcommand{\be}{\begin{equation}}
\newcommand{\ee}{\end{equation}}

\newcommand{\beq}{\begin{equation}}
\newcommand{\eeq}{\end{equation}}
\newcommand{\beqa}{\begin{eqnarray}}
\newcommand{\eeqa}{\end{eqnarray}}

\def\ba{\begin{eqnarray}}
\def\ea{\end{eqnarray}}

\newcommand{\barr}{\begin{array}}
\newcommand{\earr}{\end{array}}

\renewcommand{\S}{Section~}

\usepackage[flushleft]{threeparttable}
\usepackage{array}

\usepackage{savesym}
\savesymbol{tablenum}
\usepackage[separate-uncertainty=true, multi-part-units=single]{siunitx}
\restoresymbol{SIX}{tablenum}

\usepackage{gensymb}
\usepackage{float}
\usepackage{makecell}
\usepackage{hyperref}
\usepackage[hyphens]{url}

\providecommand{\sorthelp}[1]{}

\usepackage{natbib}

\shorttitle{}

\begin{document}

\title{Using Two-Frequency Dust Spectral Matching to Separate Galactic Synchrotron and Free-Free Temperature Foregrounds from the CMB}

\correspondingauthor{J. L. Weiland}
\email{jweilan2.jhu.edu}

\author[0000-0003-3017-3474]{J. L. Weiland}   
\affiliation{The William H. Miller III Department of Physics \& Astronomy,
Johns Hopkins University,
3400 North Charles Street \\
Baltimore, MD 21218, USA}

\author[0000-0001-8839-7206]{Charles L. Bennett}
\affiliation{The William H. Miller III Department of Physics \& Astronomy, 
Johns Hopkins University,
3400 North Charles Street \\
Baltimore, MD 21218, USA}

\author[0000-0002-2147-2248]{Graeme E. Addison}
\affiliation{The William H. Miller III Department of Physics \& Astronomy,
Johns Hopkins University,
3400 North Charles Street \\
Baltimore, MD 21218, USA}

\author[0000-0002-1760-0868]{Mark Halpern}
\affiliation{Department of Physics and Astronomy, 
University of British Columbia,  
Vancouver, BC  Canada V6T 1Z1}

\author[0000-0002-4241-8320]{Gary Hinshaw}
\affiliation{Department of Physics and Astronomy,
University of British Columbia, 
Vancouver, BC  Canada V6T 1Z1}

\begin{abstract}  
We introduce a method for removing 
CMB and anomalous microwave emission (AME, or spinning dust) intensity signals at
high to intermediate Galactic latitudes 
in temperature sky maps at frequencies roughly between 5 and 40 GHz.  The method relies on
the assumption of a spatially uniform combined
dust (AME and thermal) rms spectral energy distribution for these regions, but is otherwise model independent.
A difference map is produced from input maps at two different
frequencies in thermodynamic temperature: the two frequencies are chosen such that the rms AME signal in the
lower frequency ($\sim5-40$ GHz) map is equivalent to the thermal dust emission rms in the higher frequency 
($\sim95-230$ GHz) map.  
Given the high spatial correlation between AME and thermal dust,
the resulting difference map is dominated by synchrotron and free-free foreground components, and can thus
provide useful insight into the morphology and possible spectral variations of these components at high latitudes.
We show examples of these difference maps obtained with currently available WMAP and Planck data
and demonstrate the efficacy of CMB and dust mitigation using this method.
We also use these maps, in conjunction with Haslam 408 MHz and WHAM H$\alpha$ observations, 
to form an estimate of the diffuse synchrotron spectral index in temperature on degree scales.
The hybrid analysis approach we describe is advantageous in situations
where frequency coverage is insufficient to break spectral degeneracies between AME and synchrotron.
\end{abstract}

\keywords{ \href{http://astrothesaurus.org/uat/1146}{Observational cosmology (1146)};
\href{http://astrothesaurus.org/uat/383}{Diffuse radiation (383)};
\href{http://astrothesaurus.org/uat/856}{Interstellar synchrotron emission (856)};
\href{http://astrothesaurus.org/uat/836}{Interstellar dust (836)}
}

\section{Introduction}

Characterization of CMB foregrounds continues to be an issue of importance in cosmological analyses.
Although recent focus has been directed toward polarized foregrounds, characterizing the intensity
behavior is also relevant for understanding the physics of the Galactic emissions and best algorithms
for high precision foreground removal.  While data from the Planck mission \citep{planck/01:2018}
have empowered significant advances in the characterization of thermal dust at frequencies $\gtrsim 100$ GHz, precise separation of multiple diffuse temperature foregrounds in the WMAP and Planck LFI maps ($\sim20 - 90$ GHz, \citealt{bennett/etal:2013, planck/02:2018}) has proven difficult for several reasons.  These reasons include the number of components, 
spectral and/or spatial degeneracies between some components, and the present paucity of frequency coverage in ranges where component spectral degeneracies may be broken.

In addition to the CMB, there are four key contributors to the diffuse temperature emission in the $20 - 90$ GHz range:
synchrotron, free-free (thermal bremstrahlung), anomalous microwave emssion (AME, thought to
arise from spinning dust), and thermal dust.  While the thermal dust is most easily separable spectrally because it has a completely different frequency dependence than the other three,  it shares a strong
morphological correlation with the AME.  Similarly, synchrotron and dust have some spatial correlation,
and both synchrotron and the AME have a similar SED in a restricted frequency range at high to intermediate Galactic latitudes ($\sim \nu^{-3}$   between $20-60$ GHz).  The AME SED is predicted to vary with physical conditions and 
emitting particle type 
(e.g. \citealt{draine/lazarian:1998,ali-haimoud/etal:2009, silsbee/etal:2011,hensley/draine:2023}) , but observations of individual source SEDs in flux density units indicate a spectral shape similar to a log-normal distribution
that typically peaks near $\sim 30$ GHz  \citep{dickinson/etal:2018}. 

Physically motivated spatial templates that have consistently evidenced high correlation in a variety of AME source
studies  on $\sim$ degree scales include thermal
dust radiance (\( \mathcal{R} \), a tracer of the interstellar radiation field and dust column, 
\citealt{hensley/etal:2016})
and optical depth at 353 GHz ($\tau_{353}$, tracing HI column density,
\citealt{planck/11:2013}).
A number of observational templates of thermal dust emission in the mid- and far-IR have also been used.
However, the AME SED is clearly not representative of a thermal dust emission mechanism, and no 
individual thermal dust template has been found that adequately describes the AME emission over the entire sky.

Separation of AME, synchrotron and free-free components based on spectral shape alone can transfer power between components, due to insufficiently available frequency coverage, degeneracy between the AME and synchrotron SED in the $\sim 20-60$ GHz range, and assumed spectral models that do not sufficiently match the dependence in the data.
Examples of power transfer in 
the Planck 2015 
\textsl{Commander} intensity foreground decomposition have been noted by \cite{planck/04:2018} 
for exactly these reasons. 
Subsequent Planck releases did not revisit a full intensity component decomposition in
this complex frequency range.

An alternate separation method relies on spatial correlation of component templates with data at multiple frequencies in large
sky patches.  This method does not assume anything about component SEDs, but typically derives SEDs on $\sim10^\circ-15^\circ$
scales.  The method loses accuracy in regions where either component morphology lacks dynamic range in a given patch, or there is degeneracy between component spatial morphology.  The method also relies on 
use of an external estimate of a CMB template. 
\citet{harper/etal:2022} employ this method using data from the northern C-BASS survey (5 GHz),  WMAP (23, 33, 41, 61 and 94 GHz) and Planck LFI and HFI (30 - 857 GHz).  The addition of the 5 GHz data further constrains the AME SED, and they derive a mean AME SED for high to intermediate Galactic latitudes in the northern sky. 
Interestingly, \citet{harper/etal:2022} find their AME emissivities differ
from those obtained from SpDust2 \citep{silsbee/etal:2011} models for typical representative physical environments.

\begin{figure*}[ht]
    \centering
    \includegraphics[width=7in]{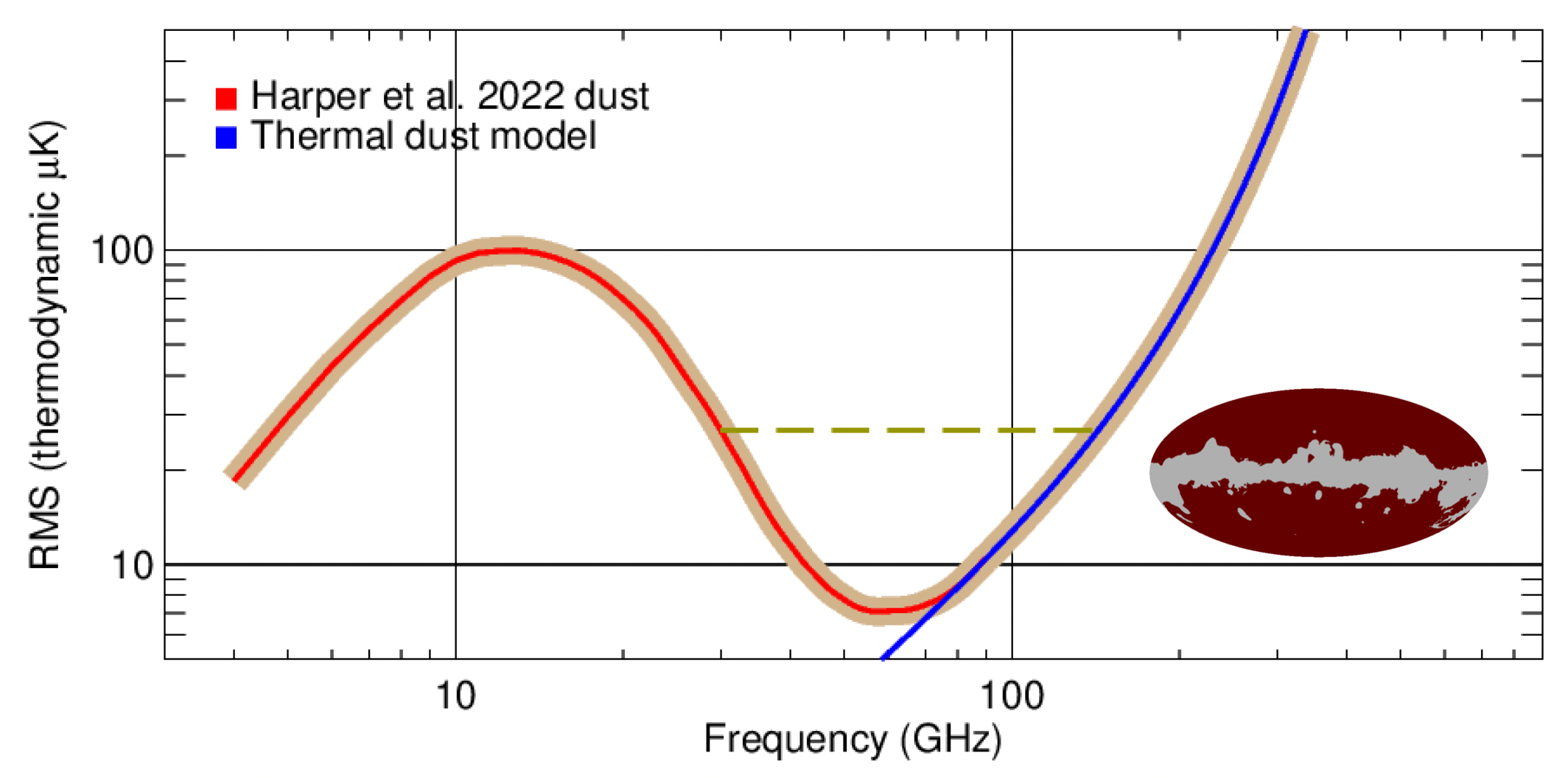}
    \caption{High to intermediate Galactic latitude AME and thermal dust emission spectrum on $1^\circ$ scales in thermodynamic units.  The blue line is the \planck\ 2015 Commander
    thermal dust model rms spectrum, evaluated over the unmasked (brown) regions of the inset mask map (shown in Galactic coordinates).  The red line is the combined AME and thermal dust rms spectrum derived by \cite{harper/etal:2022} using a template correlation method.  Since the
    \cite{harper/etal:2022} spectrum was only evaluated over the C-BASS North survey footprint, we have scaled it to the
     blue thermal dust model by matching values of the two spectra at 90 GHz.  The thick tan curve represents the combined AME and thermal dust spectrum.  Given the combined spectrum, the rms level for an AME-dominated
     frequency between $\sim$5 and 40 GHz will have a matching frequency at the same level which is dominated by
     thermal dust.  One such frequency pair is shown by the horizontal dashed green line, pairing 30~GHz (AME dominated)
     with 143 GHz (thermal dust dominated).
     In these units, the peak of the AME spectrum (between 10-20 GHz) is at a thermodynamic temperature
     comparable with the thermal dust emission at $\sim 230$ GHz.
    }
    \label{fig:rms_spec}
\end{figure*}

\begin{figure*}[ht!]
    \centering
    \includegraphics[clip,trim={0mm 3.5mm 0mm 1mm},width=6.82in]{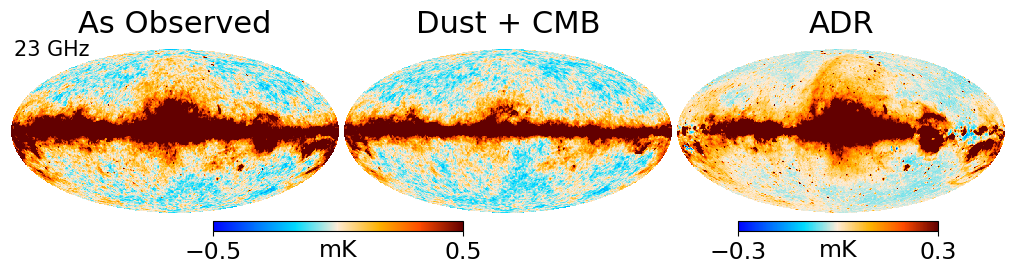}
    \includegraphics[clip,trim={0mm 3.5mm 0mm 1mm},width=6.82in]{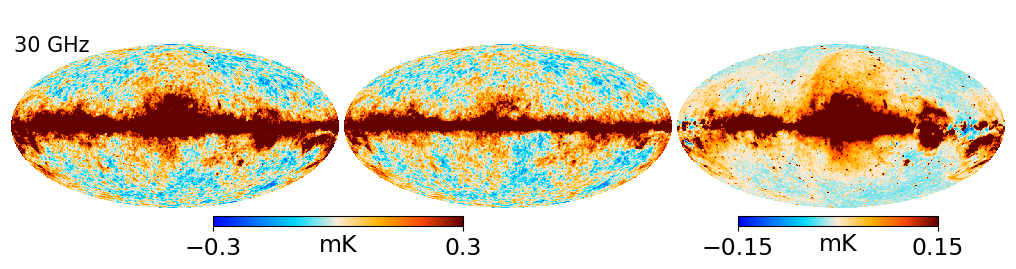}
    \includegraphics[clip,trim={0mm 3.5mm 0mm 1mm},width=6.82in]{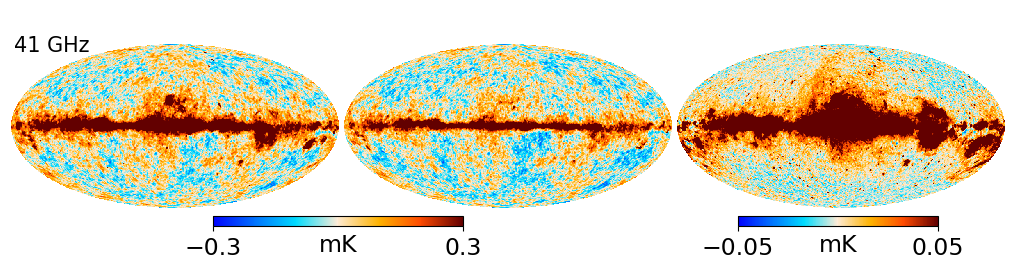}
    \caption{In the left column, as-observed WMAP or Planck maps are shown for 
    three frequencies (23, 30 and 41 GHz).
    These maps are a mix of CMB and multiple foreground signals. Monopoles have been removed to visually show a larger color contrast between CMB and foregrounds.  The middle
    column is our estimate of the combined CMB and dust signal present in the maps on the left.  The estimate is formed from a weighted combination of higher frequency maps that are dominated by CMB and thermal dust emission; weights are chosen to preserve CMB and
    conform to a specific dust SED (see text and Figure~\ref{fig:rms_spec}).
    The left and middle maps in each row form a map pair, chosen such that the difference will effectively remove CMB and high latitude AME from the left-hand map of the pair without use of a parametric fit, leaving predominantly diffuse synchrotron and free-free foregrounds.  
    The resulting difference maps (left minus middle) are shown in the right column.  We refer to these 
    difference maps as ``AME Dust Reduced'' (ADR) maps. 
   The similarity of the morphological structure between each of the three ADR frequency maps is striking. The dark brown regions close to the plane are saturated at this color stretch, and represent a complex mix of free-free, synchrotron and likely dust removal imperfections.
    Bright sources are also evident off the plane.  Outside of these regions, the dominant morphology is that of diffuse synchrotron (see Figure~\ref{fig:diff30_143}). Note that a monopole has been removed from each difference map for visual clarity.
    \label{fig:pairs_and_adr}
    }
\end{figure*}

In this paper, we use
the C-BASS observational constraint on the mean AME spectral behavior at intermediate to high Galactic latitudes,
in conjunction with linear combinations of selected frequency maps, to remove  CMB and AME 
components without 
relying on additional external templates  or modeling assumptions.  Mitigation of these two components
in low frequency maps ($\nu \lesssim 40$ GHz)
permits a closer look at the remaining synchrotron and free-free emission.
Any systematic errors in the maps are assumed to be small and are ignored in the analysis.
While the use of linear combinations of maps to nullify CMB signals or produce foreground templates is not new
(e.g., \citealt{bennett/etal:2003,bennett/etal:2013,planck/10:2015,planck/04:2018} and references therein), the particular choice described here
appears not to have been previously applied in the literature.
In Section 2. we discuss our method.  The results are presented in section 3, followed by conclusions in Section 4.

\section{Data and Separation Method}

We begin by constructing an rms dust intensity SED appropriate for high and intermediate Galactic latitudes.  
The dust SED respresents
the combined emission from the AME and thermal dust emission. Previous component separation studies \citep{planck/04:2018, planck/10:2015, harper/etal:2022} indicate fairly uniform
physical conditions in off-plane regions for both of these components, allowing us to attempt to characterize these regions with a single SED.
We generate thermal dust emission maps at frequencies between 5 and 857~GHz using the parameters from the
\planck\ 2015 \textsl{Commander} component decomposition solution for a modified black-body \citep{planck/10:2015}.
We then compute the rms thermal dust emission spectrum outside a mask that excludes regions of 
CO line emission and strong free-free emission in the Galactic plane, based on thresholded versions of the Planck CO J=1$-$0 map \citep{planck/10:2015} and WMAP 9-yr K-band MEM estimate of free-free \citep{bennett/etal:2013}.  The mask is shown as an inset in Figure~\ref{fig:rms_spec}, and the thermal dust
spectrum, converted from Rayleigh-Jeans to thermodynamic temperature units, is represented by the blue line in the plot.

A key result of the \citet{harper/etal:2022} analysis discussed in the introduction is an updated spectrum for the AME at intermediate to high Galactic latitudes.  We have digitized their rms combined AME + thermal dust spectrum from their figure~11.  
This SED was computed from maps at $1^\circ$ resolution, for high to intermediate Galactic latitudes, but only over the C-BASS North survey footprint.  We assume, however, that this mean SED is valid for all high to intermediate latitudes,
and scale it with a single multiplicative factor to match the thermal dust SED at 90~GHz in Figure~1.
This scaled trace is shown in red in that figure.  The combined AME + thermal dust SED is represented by the thick tan
curve.

\begin{table}[ht]
    \centering
    \caption{Examples of pairs of frequencies with equal amplitude AME and thermal dust RMS in thermodynamic units.}
    \label{tab:ame_tdust_freq_pairs}    
    \begin{tabular}{c|c}
    \hline
    AME Frequency  &  Thermal Dust Frequency \\
    (GHz)         &   (GHz) \\
    \hline\hline
       10   &  224 \\
       13   &  229 \\
       20   &  205 \\
       30   &  143  \\
       40   &   95  \\
     \hline  
    \end{tabular}

\end{table}

The rms dust spectrum in Figure~1 illustrates that there are AME-dominated frequencies ($\lesssim 40$ GHz) for which
there are ``companion frequencies'' in the thermal dust emission dominated regime ($\gtrsim 90$ GHz) whose rms level
is the same.  The horizontal green dashed line in the figure shows one example of such a frequency pair, and
in Table~1, further examples are given of AME-dominated frequencies (first column) for which there are thermal dust dominated frequencies (second column) whose rms values are equal in thermodynamic units (frequencies are rounded to the nearest integer).
As a specific  example, the equivalence of the AME emission at 30~GHz and that of thermal dust at 143~GHz
introduces the opportunity for simultaneous removal of CMB and AME emission from 30~GHz simply by subtracting the thermal dust dominated 143~GHz map from the 30~GHz map.  The advantage of this method is that it avoids a more complex parametric model fit to multiple frequencies
needed to solve for several emission components, and, provided the dust spectrum in Figure~\ref{fig:rms_spec} is correct, reduces the potential for power transfer between components. 
Over the regions where the spatial correlation between AME and thernal dust holds,
the residual at 30 GHz would be dominated by the remaining foreground components: a nearly undiminished 30 GHz synchrotron emission component and residual free-free emission.  The 30~GHz monopole is altered however, and the correlation is not expected not hold for sky regions within about $20^\circ$ of the Galactic plane, where a more complex and variable physical environment exists.  

The above discussion provides a guide to producing CMB and AME nulled maps from frequency pair differences,
but there are added complexities necessitated when using real data.  The closest available pairs of frequencies in public archives are those
of \Planck\ 30 GHz and 143 GHz, and WMAP 41 GHz and 94 GHz.  
The available \wmap\ and \planck\ maps do not provide measurements at 
single frequencies, but rather passbands, and the nominal frequencies of these passbands do not
exactly match those predicted to null dust.  The most deviant example is that of \WMAP\ 23~GHz, whose matching
thermal dust frequency lies between the available \planck\ 143 and 217 passbands.
Therefore, given an AME-dominated band, we construct its equivalent
thermal dust temperature map from a weighted combination of two bands that bracket the desired thermal dust frequency.  
The weights for the two thermal dust bracketing bands, $w_1$ and $w_2$, are constrained to preserve the CMB signal while producing the desired rms AME dust level:
\begin{align}
      w_1  + w_2  &  =  1 \\
      w_1 D^{rms}_1  + w_2 D_2^{rms}   &  = AME^{rms} ,
\end{align}

{\noindent}where $D^{rms}$ is the thermal dust rms in thermodynamic units computed for a band using the thermal dust model of Figure~1. Conversion factors for Rayleigh-Jeans to thermodynamic units are applied, 
and the broad-band response handled either using color corrections (for Planck bands\footnote{software and tables available from the Planck Legacy Archive.}) or
effective frequencies (for WMAP bands\footnote{\url{https://lambda.gsfc.nasa.gov/product/wmap/current/effective_freq.cgi}}).  

To produce a thermal dust map with masked rms signal matching that for the WMAP 23~GHz AME, we choose the thermal dust bracketing bands to be the
Planck 143 and 217 GHz bands and solve for the weights, yielding $w_{143} = 0.563$ and $w_{217} = 0.437$.
Simlarly, for Planck 30 GHz, the weights are $w_{143} = 0.937$ and $w_{217} = 0.063$.
Note that the \planck\ 217~GHz map also contains
emission from the J=2$-$1 CO emission line, which is not accounted for in the weight computation.  Prior to computing
the weighted linear combination of the 143 and 217 GHz maps, we subtract a J=2$-$1 CO
emission contribution from the 217 GHz band using the map of \cite{ghosh/etal:2023}, converted to thermodynamic temperature units 
\footnote{\url{https://wiki.cosmos.esa.int/planckpla/index.php/UC_CC_Tables##CO_unit_conversion}},
smoothed to $1^\circ$ FWHM and thresholded at 3~$\mu$K to further reduce the noise component at high latitudes. 
Residual contributions from the Sunyaev-Zeldovich (SZ) effect are 
subdominant compared to the diffuse Galactic foreground signal strength on degree scales and are ignored in our analysis (see e.g., \citealt{planck/10:2015} and Planck FFP10 sky model).

In the left and middle columns of Figure~\ref{fig:pairs_and_adr}, we show three map pairs before differencing: \wmap\ 23 GHz and its [143, 217 GHz] weighted companion map, \planck\ 30~GHz and its [143, 217 GHz] companion map (which is dominated by the 143 GHz contribution), and \wmap\ 41 GHz and its 94 GHz companion map.  
Maps are in thermodynamic temperature units and have been smoothed to a common resolution of $1^\circ$ FWHM and pixelized at HEALPix \citep{gorski/etal:2005}
resolution $N_{side}=128$.
The \wmap\ data are the full mission maps from the 9yr data release \citep{bennett/etal:2013}, and the \planck\ data are full mission frequency maps 
from the 2018 PR3 release \citep{planck/01:2018}.  
At this exploratory level, we chose to use the PR3 release over that of PR4 \citep{npipe:2020}: use of PR3 
avoids the  
necessary additional step of removing the intentional zodiacal light residual in the PR4 intensity maps.
To ensure consistent treatment of the kinematic quadrupole between \wmap\ and \planck\ maps, we have subtracted  this component from the 23 GHz map.  
We do not produce an exact companion map for 41 GHz because
the 94~GHz map is within a few percent of the necessary rms level, and a [70, 94 GHz] weighted combination contains a more complex mix of foreground components than the weighted [143, 217 GHz] maps.  In addition to dust emission, much of the shared higher latitude visible morphology between the maps in the left and middle columns of the figure is from the CMB.  This is particularly true for the 41 and 94 GHz map pair, because
the foregrounds are relatively weaker.

We discuss the map pair residuals and how we interpret them in the following section.

\section{Results}

Differences between the map pairs of Figure~\ref{fig:pairs_and_adr} are shown in the right
column of that figure.
By construction, the differencing operation nulls (to within calibration uncertainties) the CMB component that was present in both maps of each pair, and the remaining sky signal in the difference map is attributable to foregrounds.
We refer to the pair difference maps as ``AME Dust Reduced'' (ADR) maps, in which the AME component in the
lower frequency of the map pair has been mitigated by subtracting the thermal dust emission present in the higher frequency companion map.

A frequency-dependent monopole has been removed from each of the difference maps in Figure~\ref{fig:pairs_and_adr}, and color stretch maxima scaled assuming $\sim \nu^{-3}$, to assist with displaying
the similarity in structure. 
Visually, the morphologies of all three ADR maps 
are strikingly similar.  
This in itself provides an important check of the validity of the assumed dust SED and methodology.
The common-mode morphology can be interpreted as an indicator of successful AME mitigation with frequency, and we therefore proceed to examine the ADR map characteristics in further detail.
For the remainder of the discussion, we concentrate
on the 23 and 30 GHz ADR maps: as stated in the previous section, the component separation in the 41 GHz ADR map is more uncertain, and its S/N at high Galactic latitudes is low.

\begin{figure}[ht]
    \centering
    \includegraphics[clip,trim={2mm 5mm 2mm 5mm},width=3.5in]{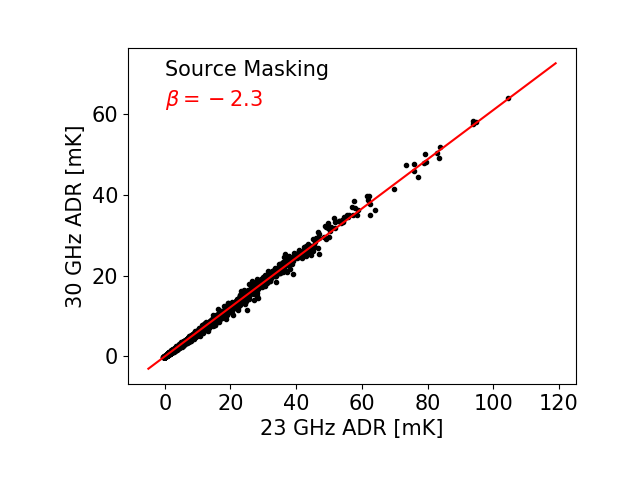}
    \includegraphics[clip,trim={2mm 5mm 2mm 8mm},width=3.5in]{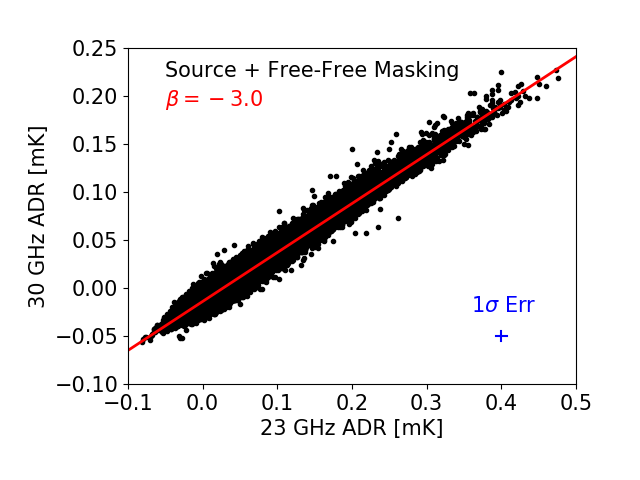}
    \caption{ Correlation plots of the 23 GHz and 30 GHz ADR maps are shown for each of two different masks.  At top,
    only pixels at locations of bright sources are excluded.  The correlation slope (red line) corresponds most closely to that of free-free emission, with a spectral index $\beta$ near $-2.3$. The bottom plot shows the correlation after both
    bright sources and strong free-free emission regions are excluded (see the mask in 
    Figure~\ref{fig:diff30_143}).  Not only is the dynamic range
    of the plot reduced, but the correlation slope corresponds to $\beta \approx  -3$, more typical of synchrotron.
    The blue cross at bottom right indicates typical statistical uncertainties per pixel (invisible in the top panel).
    Note that the two ADR maps are not entirely
    statistically independent, since they share 143 and 217 data, albeit with different weights.  However, the
    scatter about the fit line suggests some spectral variability at high and intermediate latitudes.
    }
    \label{fig:tt_plots}
\end{figure}

The top panel of Figure~\ref{fig:tt_plots} 
shows the tight correlation between the 23 and 30 GHz ADR maps
after masking of bright point sources. On the linear scale shown, the correlation is dominated by the higher signal pixels 
found concentrated toward the plane and occur primarily within the   
dark brown regions that saturate at the color table stretch
chosen for the right column of Figure~\ref{fig:pairs_and_adr}.  These are also the regions where the component decomposition is least well characterized.
At minimum, these high signal regions will contain a mix of free-free and synchrotron emission. However, it is likely that significant
dust-related residuals exist here as well, since the key supposition of our dust mitigation technique is that of shared and
fairly invariant physical conditions -- an assumption which is expected to be invalid for complex regions
closer to the Galactic plane.

\begin{figure}[htbp]
    \centering
    \includegraphics[clip,trim={0mm 1mm 0mm 12mm},width=3.20in]{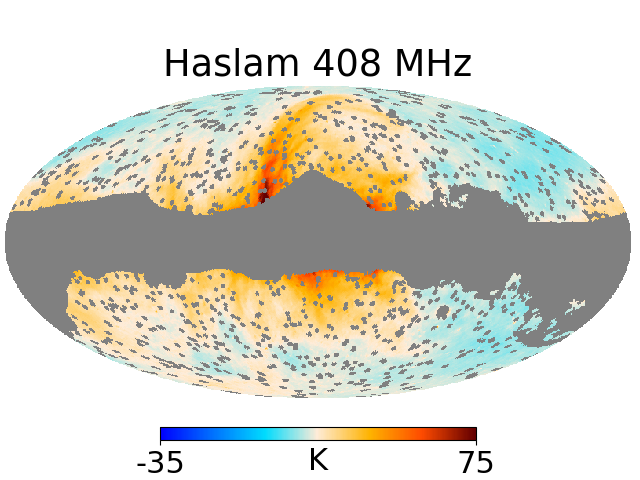}
     \includegraphics[clip,trim={0mm 1mm 0mm 5mm},width=3.20in]{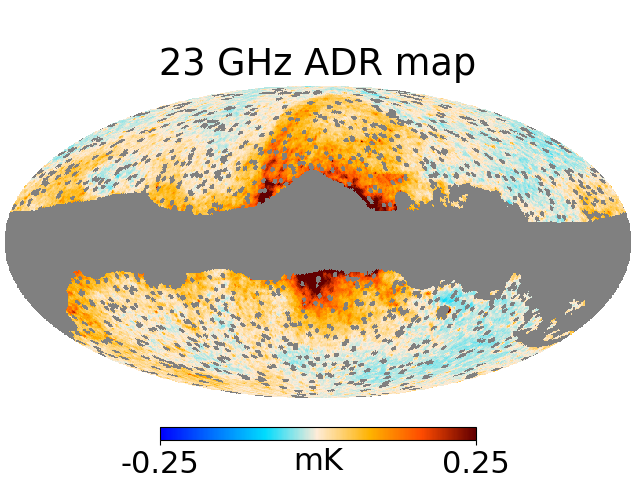}
     \includegraphics[clip,trim={0mm 1mm 0mm 5mm},width=3.20in]{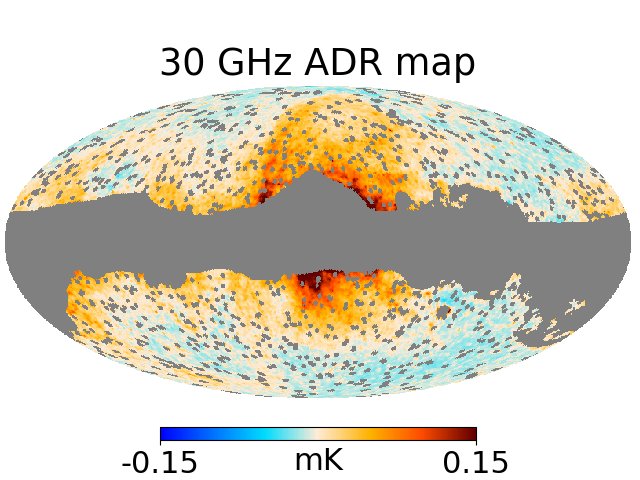}
      \caption{The Haslam map (top, as reprocessed by  \cite{remazeilles/etal:2015}), is generally recognized
      as a template for Galactic synchrotron emission at 408 MHz, especially outside the masked region that contains sub-dominant free-free emission.
      Compare this synchrotron morphology with the spatial appearance of
      the 23 GHz (middle) and 30 GHz (bottom) ADR maps for a masked portion of the sky.  
      This synchrotron morphology is seen to share the patterns in the 23 GHz (middle) and 30 GHz (bottom) ADR maps.
      The mask (gray regions in all panels)
      excludes bright sources and regions of high residual free-free emission in the ADR maps.
     The ADR methodology is designed to remove CMB and nearly all the dust-correlated AME from the 23 and 30 GHz maps with only a small and quantifiable loss of synchroton power.
     The similarity of the Haslam and ADR maps, as well as the similarity of the two ADR maps, indicates the effectiveness of the AME removal. 
     See text and Figure~\ref{fig:cockroach_plots} for further detail about the correlation.}
    \label{fig:diff30_143}
\end{figure}

We construct a sky mask (``the analysis mask'') to exclude these more complex regions from analysis and examine the higher latitude sky regions
where we expect physical conditions to be more homogeneous.  This mask is similar to the mask used in Figure 1 but does not explicitly avoid CO clouds and has a smoother diffuse boundary with fewer ``islands''.
The diffuse mask is formed
from the union of two masks: (1) the \planck\ 2015 HFI Galactic plane mask that leaves 70\% of the sky 
\footnote{\url{HFI_Mask_GalPlane-apo0_2048_R2.00.fits, field GAL070}} and (2) a free-free mask based on the WHAM 
(Wisconsin H-Alpha Mapper, \citealt{haffner/etal:2003, whamsouth:2010}) H$\alpha$
full-sky survey\footnote{\url{https://lambda.gsfc.nasa.gov/product/foreground/fg_wham_h_alpha_map_get.html}} which has been de-extincted using the method of \cite{bennett/etal:2013} and 
thresholded at 5 Rayleighs.  A bright source mask is also included, formed from the union of \wmap\ and \planck\ 30
GHz point source catalogs and augmented to include sources of larger extent such as Cen A, Fornax A, the LMC and the SMC.

\begin{figure}[ht]
     \includegraphics[clip,trim={0mm 2mm 0mm 8mm},width=3.25in]{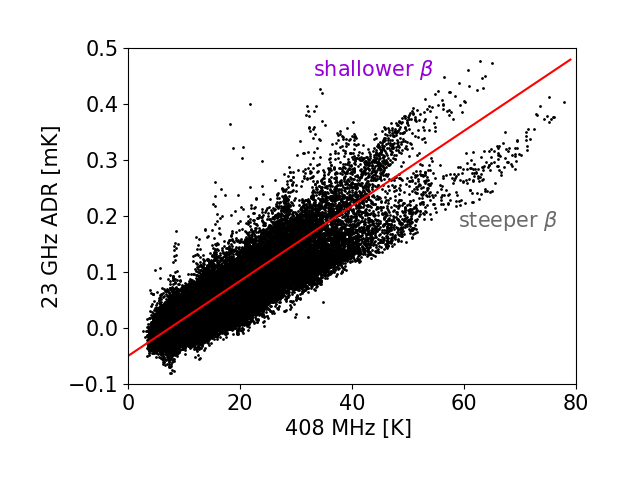}
     \includegraphics[clip,trim={0mm 2mm 0mm 8mm},width=3.25in]{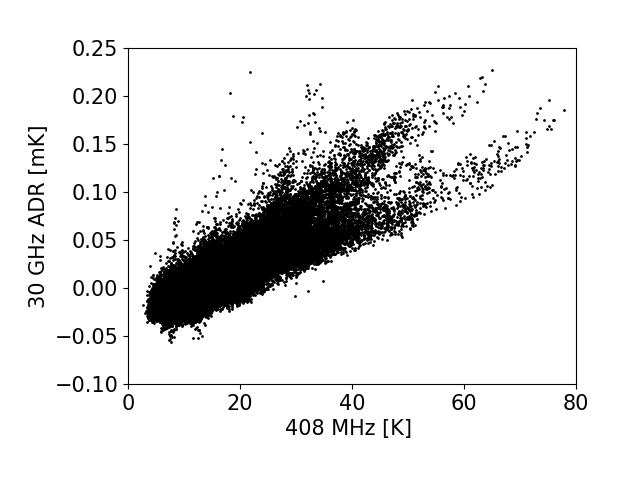}
     \includegraphics[clip,trim={0mm 2mm 0mm 10mm},width=3.25in]{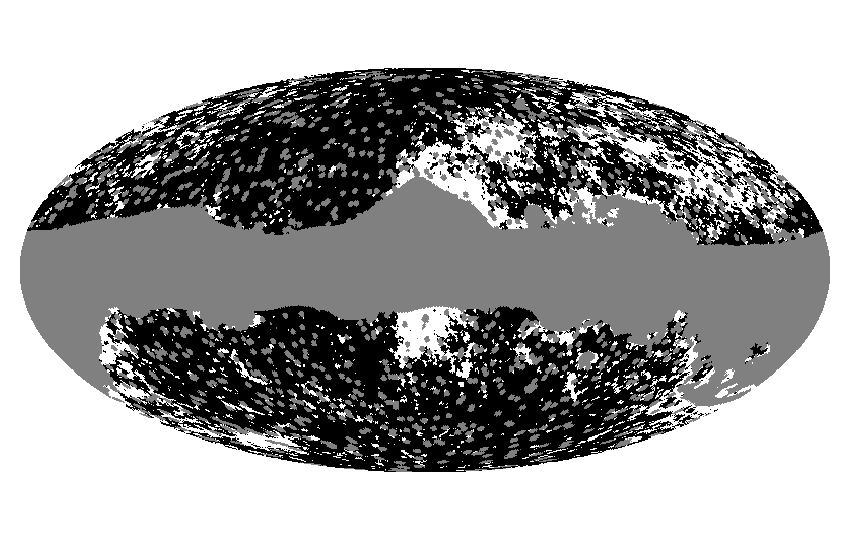}
     \caption{The correlation between the Haslam 408 MHz synchrotron map and either the 23 GHz ADR (top) or 30 GHz ADR map (middle) for pixels at high to intermediate Galactic latitudes.  The correlations show
     a bimodal distribution, indicating two regimes of spectral index $\beta$.  
     The red line in the top plot is visually estimated to separate the two regimes.
     In the bottom map, white pixels correspond to locations in the 23 GHz ADR map where the temperatures lie above the red line, representative of the shallower of the two index regimes. Gray pixels are excluded from the correlation and all other pixels are black. This appears to show genuine
     spectral index variantions, including the shallower "Haze" about the Galactic center.
     (The spectral index map in Figure~\ref{fig:raw_beta} shows corresponding values.)}
     \label{fig:cockroach_plots}
\end{figure}

Masked versions of the 23 GHz and 30 GHz ADR maps are shown in Figure~\ref{fig:diff30_143}, along with the equivalently masked version of the Haslam 408 MHz survey, which is dominated by synchrotron emission.  Visually, there is a strong correlation
between Haslam and these two maps, which also argues for successful mitigation of the AME component. However, correlation plots between masked pixels in the Haslam and ADR maps 
(top and middle panels in Figure~\ref{fig:cockroach_plots} ) show
a bimodal distribution with two distinct ``tails'': the upper representing a steeper slope, indicative of a shallower spectral index, and a lower tail representing a steeper spectral index.  
The intercepts of the correlation plots are somewhat arbitrary, since monopoles for the ADR maps have not been adjusted to Galactic levels.

We use a thresholding method to isolate the approximate spatial location of the upper tail (shallower index) points.  The red line in the top of plot of Figure~\ref{fig:cockroach_plots} divides the bimodal distribution into two portions.  This line is an approximate divider,  determined solely from visual inspection
of the apparent fork in the point distribution. 
Pixels in the 23 GHz ADR map where the temperatures lie above the red line are shown in white in the bottom map of Figure~\ref{fig:cockroach_plots}.  Gray pixels are excluded from the correlation and all other pixels are black.  
While the steeper spectral index regime in black is likely representative of high latitude synchrotron emission
with sub-dominant free-free emission, the shallower index regime in white is more difficult to categorize.
The shallower index may indicate synchrotron spectral variations, or a mix of bright free-free and synchrotron, or for some isolated locations, may indicate the presence of residual AME.  The extended white region in the bottom panel of Figure~\ref{fig:cockroach_plots} is likely associated with the diffuse shallower index
``Galactic Haze'' surrounding the Galactic center \citep{finkbeinerhaze:2004, pietrobon/etal:2012, planck/intermediate/09:2013}.  

Additional clarification of the index spatial distribution can be gained by examining
the full-sky spectral index at one-degree resolution using the Haslam 408 GHz map in conjunction with the 23 GHz and 30 GHz ADR maps. Computation of an index map requires that Galactic emission monopole levels
be established for the maps.  For the Haslam map, we adopt the estimate of the excess extragalactic emission used in the Planck 2015 analysis \citep{planck/10:2015,wehus/etal:2014}.  
Offsets for the 23  and 30 GHz maps are computed using a template-based method that is described
in further detail in Appendix~\ref{app:monopoles}.
The monopole adjusted maps are produced by subtracting
offsets of 8.9 K, $-$0.052 mK and $-$0.042 mK from the 0.408, 23 and 30 GHz maps.  A power law is fit to these three frequencies per
$N_{side}=128$ pixel, constrained such that the 408 MHz data point is fixed: $T_{[\nu=408,23,30]} = T_{408} (\nu/0.408)^\beta$ in antenna temperature, with $\nu$ in GHz.  Note that because we are modeling (at least) two components as a single power law, this is not a completely accurate model of the emission.

\begin{figure}
    \centering
    \includegraphics[clip,width=3.45in]{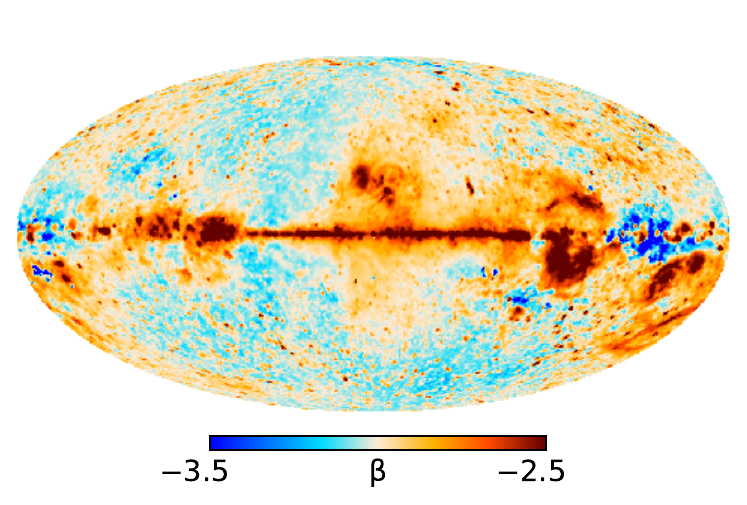}
    \caption{A map of the spectral index resulting from a single power law fit per pixel using Haslam 408 MHz, 23 GHz ADR and 30 GHz ADR maps
    smoothed to $1^\circ$ resolution and pixelized at HEALPix $N_{side}=128$. As expected, there is a general correspondence between
    the shallower index (orange shades) regions with the white areas of the bottom map of Figure~\ref{fig:cockroach_plots}.
    A single power law is not an ideal model, since the mixture of foregrounds is represented by at least two power laws of different spectral 
    indices (i.e., synchrotron and free-free). Steep power law indices (dark blue) near the anticenter are not expected if the plane emission is dominated
    by free-free.  We interpret this as an indicator of potential systematic residuals in the ADR maps near the plane, where physical conditions likely do not conform with the uniform dust SED assumption adopted for higher latitudes.
    }
    \label{fig:raw_beta}
\end{figure}

The single power-law fit is shown in Figure~\ref{fig:raw_beta}.  Although the fit is not intended for quantitative study, it is useful for illustrating two points: (1) it supports the interpretation of the two-tailed distribution in the correlation plots as divided into two index regimes (orange and blue) at 
intermediate and high latitudes, and (2) it suggests the presence of a systematic
error in the ADR maps near the Galactic plane.  This latter point is based on the
presence of steep (blue color) $\beta$ values near the plane anticenter, whereas the expectation of a free-free
dominated plane would show a more uniform orange or brown coloration.  We note these areas correspond to
regions of possible signal oversubtraction in the ADR maps of Figure~\ref{fig:pairs_and_adr}.
We next attempt to provide a component separated spectral index analysis.

\begin{figure} [t]
    \centering
    \includegraphics[clip,trim={2mm 2mm 0mm 5mm}, width=3.25in]{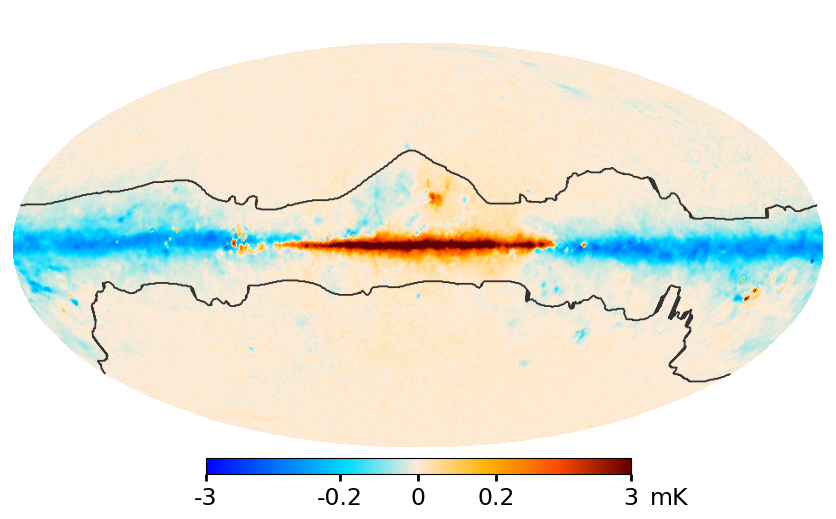}
    \caption{ ADR maps are constructed under the assumption of a spatially uniform dust SED, a premise that
    is likely invalid near the Galactic plane.  
    We use a template method to estimate the systematic bias introduced at low latitudes in the ADR maps as a consequence of this assumption (see text).
    The image shows our estimate of the bias for the 23 GHz ADR map.  The pattern
    indicates an oversubtraction of dust signal in the outer plane and an undersubtraction in the inner plane.
    The color stretch is 
    linear between $\pm 0.2$ mK, but logarithmic beyond that to show more detail close to the plane.
    Positive  values beyond 3 mK are clipped in the image.  
    The bias estimated for the 30 GHz ADR map is  of similar morphology.  Although we show full-sky
    residuals, the bias estimate is intended only for pixels interior to the analysis mask, indicated by the black contours. 
    We subtract this bias estimate from ADR map pixels interior to the mask prior to performing our
    (uncertain) determination of the 
    synchrotron spectral index over the entire sky.}
    \label{fig:fgbias}
\end{figure}

\subsection{Estimated Synchrotron Spectral Index Map}

In this section, we derive an estimated all-sky synchrotron spectral index map using only the 408 MHz Haslam map and the 23 and 30 GHz ADR maps.  The use of only three frequencies requires some assumptions, and the choices we make are motivated in part by the wish to avoid direct use of component templates that have been derived from previous parameterized decomposition efforts.  
Ideally, we would perform a new decomposition utilizing as many frequencies as possible, but defer that
to future work, as the primary purpose of this paper is to illustrate the utility of the ADR concept. 
Given these constraints, we employ two analysis assumptions: 
(1) we assume we have an independent estimate of the free-free emission at all three frequencies, and (2) we assume we have an estimate of the foreground subtraction bias introduced at low Galactic latitudes when we created the 23 and 30 GHz ADR maps.

In Appendix~\ref{app:monopoles}, we describe our estimated free-free template at 23 ~GHz based on an extinction-corrected version of the all-sky velocity integrated WHAM H$\alpha$ map\footnote{\url{https://lambda.gsfc.nasa.gov/product/foreground/fg_wham_h_alpha_map_info.html}}.
We scale the extinction corrected H$\alpha$ map to units of mK at 23~GHz from its original units of Rayleighs by 
using a factor of 7.4 $\mu$K/R.  As discussed in the Appendix, this value is consistent with previously published results.
We then extend the free-free estimate to 408 MHz and 30 GHz by assuming power-law behavior, with
a spectral index of $-2.11$ between 23 and 408, and $-2.14$ between 23 and 30 \citep{bennett/etal:2003, planck/10:2015}.

We estimate the foreground subtraction bias introduced in the ADR maps at low Galactic latitudes by computing the difference between a scaled spatial template for AME and the ADR thermal dust companion maps.  
Recently, \cite{quijote_plane:2023} used a parametric fit to QUIJOTE MFI $11 - 19$~GHz northern sky observations in conjunction with multiple other datasets to derive a model of AME emission over much of the Galactic plane ($|b| \le 10^\circ$).  They found a high correlation between the COBE/DIRBE \citep{hauser/etal:1998}
240 $\mu$m dust map and 
their derived AME amplitude at peak emission.  While there is no
single existing mid- or far-IR dust template that has been shown to perfectly describe the AME emission over the full sky,
we choose to use the DIRBE 240 $\mu$m map\footnote{\url{http://cade.irap.omp.eu/documents/Ancillary/DIRBE_ZSMA/DIRBE_ZSMA_10_1_256.fits}}
as an approximate template for the spatial morphology of AME based on the
QUIJOTE results.   

To produce the frequency-dependent ADR bias correction map, we perform a linear fit  that scales the
AME template ($D_{240}$) emission to that of the CMB-subtracted ADR thermal dust companion map ($D^{ADR}$):
\begin{equation}
D^{ADR} = aD_{240} + c.
\end{equation}
The Planck PR3 full mission SMICA 
map\footnote{\url{COM_CMB_IQU-smica_2048_R3.00_full.fits}} is used for the CMB estimate.
The fit is performed only for pixels within the analysis mask (non-gray pixels in Figure~\ref{fig:diff30_143}).

By construction, the residuals are minimized in the fitting region,
leaving an estimate of the bias at low Galactic latitudes that were excluded from the fit. An image of the estimated introduced bias for the 23~GHz ADR map is shown in 
Figure~\ref{fig:fgbias}.  The bias is computed separately for the 23 and 30~GHz ADR maps, since the 143 and 217 GHz
maps are weighted differently. 
The bias estimate predicts an oversubtraction of dust signal in the outer plane of the ADR map, in agreement with empirical evidence for this effect noted in the previous section.
In practice, there is some free-free which is absorbed into the bias map, and although the contribution is small, we account for it in our free-free removal treatment.

\begin{figure}[t]
    \centering
    \includegraphics[clip,trim={0mm 2mm 0mm 8mm},width=3.25in]{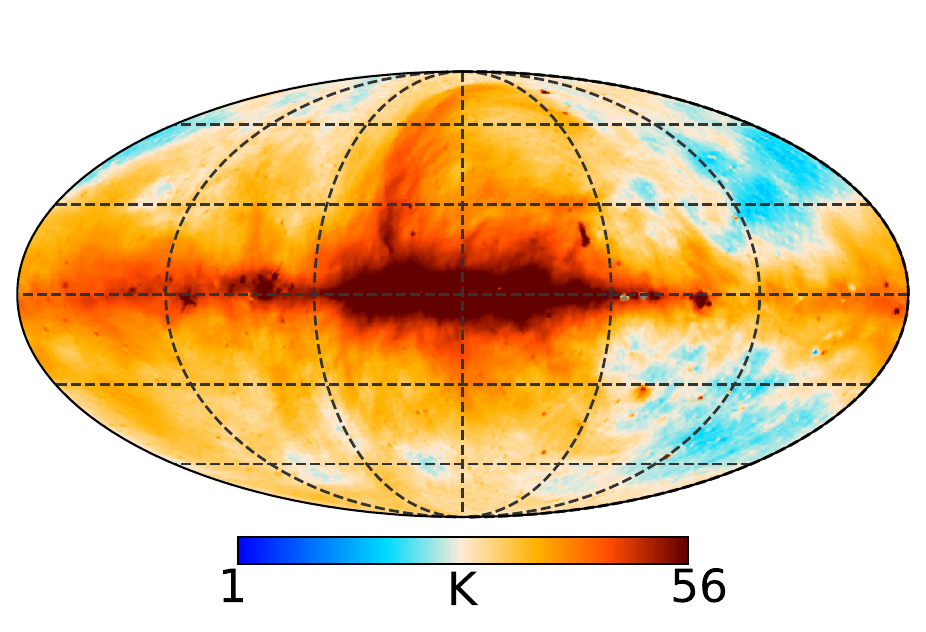}
    \includegraphics[clip,trim={0mm 0mm 0mm 4mm},width=3.25in]{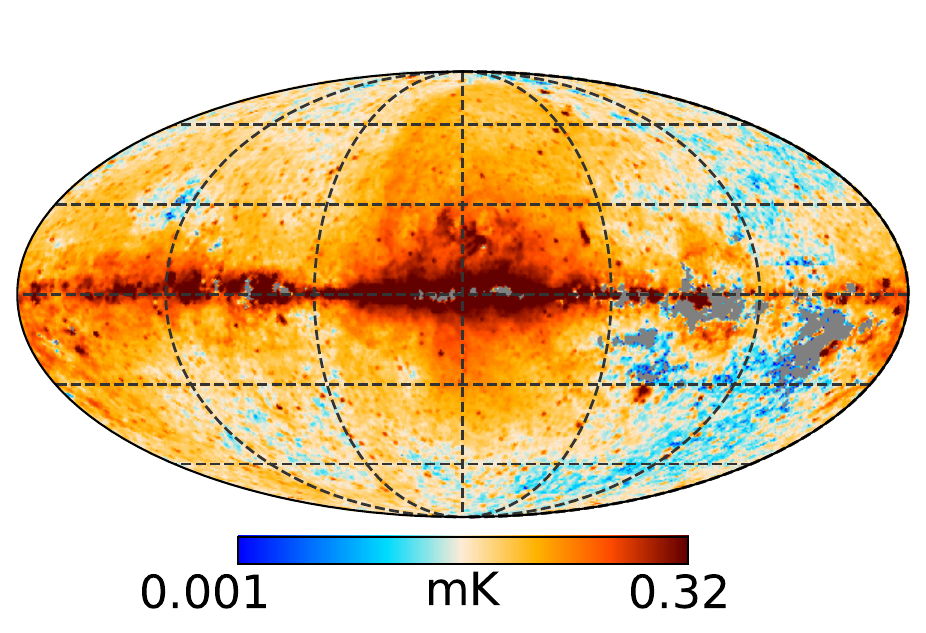}
    \caption{The 408 MHz and the bias-corrected 30 GHz ADR maps are shown after subtraction of an estimate of free-free,
    at top and bottom respectively. The scales are logarithmic. Grid lines are in Galactic coordinates with
    $\Delta l= 60^\circ$ and $\Delta b = 30^\circ$. Regions with negative temperature after free-free removal
    are shown in gray.  These two maps, together with the equivalently treated 23 GHz ADR map, serve as the input maps for a
    per-pixel power law fit that we use to estimate the synchrotron spectral index $\beta_s$. }
    \label{fig:p30_synch}
\end{figure}

\begin{figure}[th]
    \centering
    \includegraphics[clip,trim={0mm 2mm 0mm 8mm},width=3.25in]{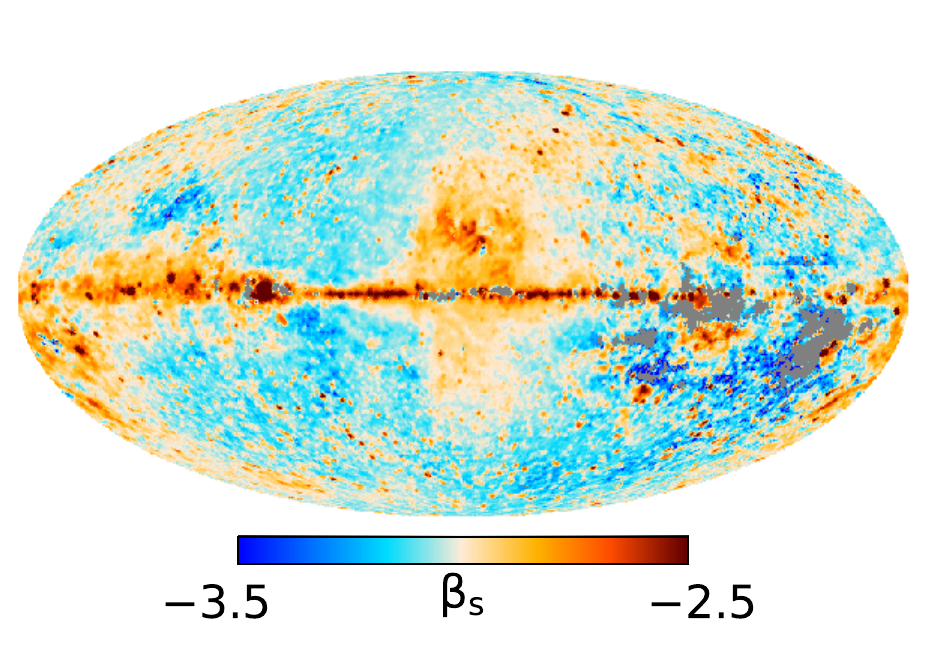}
    \includegraphics[clip,trim={0mm 2mm 0mm 6mm},width=3.25in]{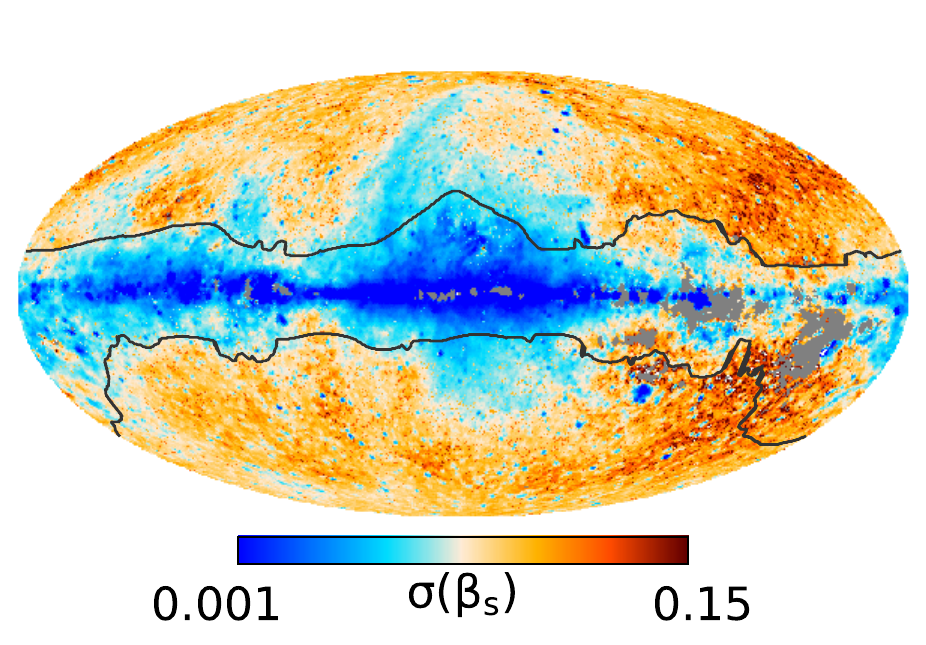}
       \caption{Top: An estimate of the synchrotron spectral index in temperature after informed, but uncertain, choices.
    The $\beta_s$ map is computed from  0.408, 23 and 30 GHz input maps (see Figure~\ref{fig:p30_synch}). Gray regions indicate oversubtraction of the free-free template in the input maps. Sources have not been removed from the map.
    Bottom: Statistical uncertainty of the power law fit to the three input maps, on a logarithmic scale. Interior to the mask outlined in black, the uncertainty is dominated by systematic, rather than statistical, uncertainties.}
    \label{fig:synch_beta}
\end{figure}

Finally, the  synchrotron spectral index map is formed from a least-squares fit of a synchrotron 
power-law model
($T_\nu = A_{408} (\nu/0.408)^{\beta_s}$  to each pixel in the three input frequency maps: 
\begin{align}
      T_{408} - FF_{408} \\
      ADR_{23} - corr23 - FF_{23} \\
      ADR_{30} - corr30 - FF_{30},
\end{align}
\noindent{where $corr23$ and $corr30$ are the foreground bias corrections to the ADR maps at 23 and 30 GHz, and
$FF$ is the estimated free-free for each frequency.}  
The free-free estimates for the ADR maps account for free-free signal loss incurred from application of the bias correction
discussed above and also any free-free loss incurred in the creation of the ADR maps themselves, i.e., 
free-free removed when subtracting the thermal dust companion maps. 
The input maps for 408 MHz and 30 GHz are shown in Figure~\ref{fig:p30_synch}; the 23 GHz map has a similar appearance to the 30 GHz map.  The maps are shown on a logarithmic scale to enhance lower signal features.
Grayed areas in the 30 GHz (and 23 GHz) maps located along the inner Galactic plane and on the left-hand side near the Gum and Orion regions ($ 210^\circ \lesssim l \lesssim 280^\circ$)
indicate pixels with negative values after the free-free removal.  These represent pixels where the assumptions
used to form the synchrotron component maps are suspect, and are excluded from analysis.  While the grayed regions
are over-subtractions, there are also indications in the Gum region ($\ell \sim 270^\circ$) of positive residual free-free correlated features.  Although we do not exclude these pixels from the fit, we also regard them as suspect.

Results of the fit are shown in Figure~\ref{fig:synch_beta}, with the spectral index $\beta_s$ shown at top and 
the statistical uncertainties at the bottom.  Interior to the mask outlined in black in the bottom plot,
systematic uncertainties dominate the error budget and the statistical portion alone does not provide an adequate characterization of the uncertainties.  As examples of systematics in this region, we have already noted pixels of negative temperature in
the 23 and 30 GHz ADR synchrotron maps, as well as areas of possible free-free undersubtraction in Figure~\ref{fig:p30_synch}.  Additionally, there is a ridge of shallow $\beta_s$ ($> -2.5$) along the inner
Galactic plane in Figure~\ref{fig:synch_beta}, co-located with pixels for which 
there is a high density of Planck 30 GHz point sources \citep{planck/04:2018} and where both the ADR bias 
and H$\alpha$ extinction corrections are largest.
As a particularly complex region, it is unlikely to be reliably fit using the templates employed here and thus we identify it as likely subject to a systematic bias.

\begin{figure}[t]
    \centering
    \includegraphics[clip,trim={10mm 0mm 0mm 0mm},width=3.5in]{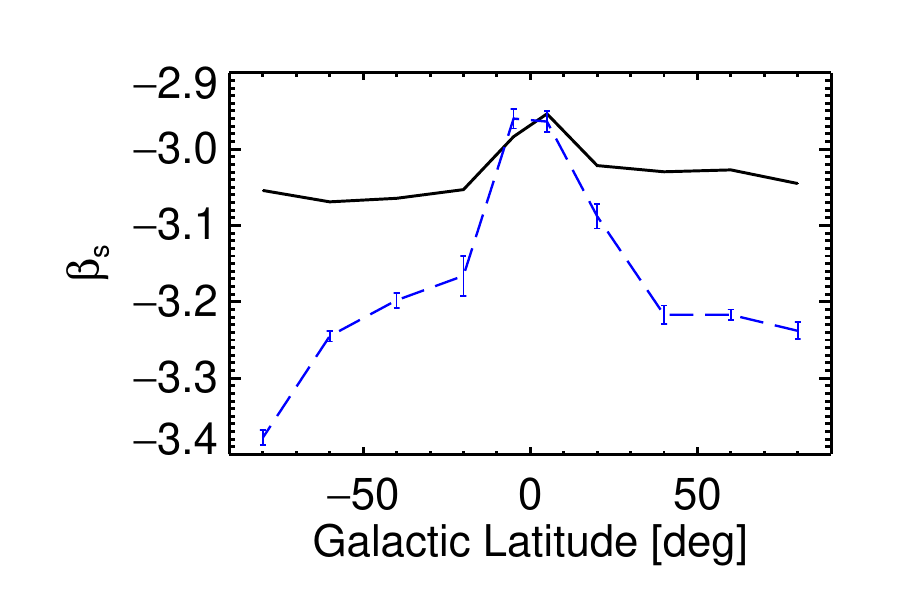}
    \caption{
    Mean Galactic latitude dependence of $\beta_s$ as derived from binning spectral index maps
    from intensity (black) and from polarization (dashed blue). 
    The profile in black is from the intensity data presented in this paper.
    Point sources and regions of no confidence, described in the text, were masked. 
    The dashed blue profile is taken directly from the polarization data in Table 2 of
    \cite{weiland/etal:2022}.  Error bars
    represent statistical uncertainties only. 
    The black trace error bars are too small to be seen
    given the y-axis scale.  Both determinations are 
    highly dependent on the fidelity of the lowest frequency map used in each analysis. No
    curvature or steepening with frequency has been included in these analyses.
    With these caveats, we conclude 
    that the intensity map is not a good predictor of
    polarization values at intermediate to high latitudes.
    }
    \label{fig:lat_profiles}
\end{figure}

In Figure~\ref{fig:lat_profiles}, we compare the mean Galactic latitude profile of our estimated intensity $\beta_s$ map with that tabulated in
Table~2 of \cite{weiland/etal:2022} for polarization.  As with previous investigations that combine
Haslam data with Planck and/or WMAP frequencies to determine $\beta_s$, we find typical high latitude 
intensity values that are shallower than those determined for polarization data by $\Delta\beta_s \sim 0.1 - 0.2$
(e.g. 
\citealt{harper/etal:2022, thorne/etal:2016, weiland/etal:2022, krachmalnicoff/etal:2018jp, fuskeland/etal:2021}).   Curvature and/or steepening of the synchrotron spectrum with frequency could impact this
result, but current information about this aspect is limited.

\section{Summary and Conclusions}

The apparent low variance in ISM physical conditions at high and intermediate Galactic latitudes, coupled with the similarity between AME and thermal dust morphology, motivate us to explore a map-based method for separating AME and CMB emission from synchrotron and free-free components.  This paper introduces a method that uses the combined AME and thermal dust temperature SED determined by \cite{harper/etal:2022} to predict frequency pairs which share an 
equivalent dust emission rms in thermodynamic temperature over a large fraction of the sky.
The pair consists of a low frequency for which AME is the dominant dust component, and a higher companion frequency for which thermal dust emission dominates.   Subtraction of the higher frequency
from the lower one results in a map dominated by the lower frequency synchrotron and free-free emission, with 
substantially mitigated dust emission and CMB nulled within calibration uncertainties.

We refer to the resultant difference map as an ``AME dust reduced'' (ADR) map at the lower frequency.
The methodology can be used for any temperature map between roughly 5-40 GHz, given complementary observations dominated by thermal dust foreground emission at frequencies between 90 - 230 GHz.  In practice, lacking an exact frequency match for the necessary thermal dust companion map, a linear combination of two bracketting high frequency maps weighted to preserve both the CMB and thermal dust spectrum may be used.  

We demonstrate the effectiveness of the ADR technique using a worked example applied to the WMAP 23 GHz and Planck 30 GHz temperature maps, both of which have high levels of AME, synchrotron and free-free emission that overlay the
underlying CMB signal.  We compute the companion (dust + CMB) map appropriate to each of these frequency bands using a linear combination of 143 GHz and 217 GHz Planck HFI maps, and subtract them to produce ADR maps at 23 and 30 GHz. 
Our main findings are as follows:

\setlist[enumerate]{label={(\arabic*)}}
\begin{enumerate}
    \item {The 23 and 30 GHz ADR maps are strikingly similar in appearance, with a high correlation between the two. This similarity may be taken as a verification of the two-frequency method. Also, importantly, this provides evidence that the AME SED is approximately consistent with the \cite{harper/etal:2022} spectrum across much of the sky.}

    \item{For a masked region that excludes sources as well as strong free-free and potential dust residuals near the plane,
    the ADR maps correlate well with the Haslam 408 MHz synchrotron template, but show a bimodal distribution
    indicative of synchrotron spectral variations and/or localized regions of bright free-free mixed with synchrotron.  
    }
    \item{Use of companion maps formed from effective frequencies $\gtrsim 143$~GHz preserves the synchrotron signal in the 23 and 30 GHz ADR maps.
    There is frequency-dependent loss of up to a few percent (maximum of $\sim 4$\% at 30 GHz) for the free-free component, due to the presence of non-negligible free-free in the companion maps.  This is a consequence of the flatter free-free spectral index.  For our worked example, the statistical noise of the ADR maps is 
    dominated by that contributed from the 23 and 30 GHz maps.} 

    \item{The assumption of uniform dust emission properties breaks down at lower Galactic latitudes.
    We estimate frequency-dependent templates of the bias introduced in the ADR maps as a result.  The
    morphology of the bias templates is similar in nature to the thermal dust temperature gradient 
    seen in far-infrared and microwave data (e.g., \citealt{planck/10:2015, sfd:1998}).
    We form ``bias corrected'' ADR maps by subtracting these estimated templates within a limited sky region; higher Galactic latitudes in the ADR maps are left uncorrected.}

\item{We remove an estimate of the residual free-free emission in the 
23 and 30 GHz bias-corrected ADR maps based on an extinction corrected WHAM H$\alpha$ template.
This last step creates estimated 23 and 30 GHz synchrotron maps.  In conjunction with the
Haslam 408 MHz map, we perform a pixel-by-pixel power-law spectrum fit to these three frequencies,
deriving an estimated synchrotron intensity spectral index on degree scales.  The baseline process
is not intended for application at lower Galactic latitudes, but we include an estimate in this region
for completeness in sky coverage, and define a mask for which we believe the results to be largely unbiased.}

\item{The resulting $\beta_s$ map has a shallower index than that reported for polarization at intermediate and
high Galactic latitudes, but both share morphology that may be associated with the Galactic Haze and 
Fermi bubbles/eROSITA lobes \citep{fermibubbles0:2010,fermibubbles:2010,carretti/etal:2013,krachmalnicoff/etal:2018jp,erosita:2020}. 
Values of the index presented in this paper depend heavily on the fidelity and calibration of
the Haslam 408 MHz map;  use of additional high quality observations at frequencies $0.4 \lesssim \nu \lesssim 3$ GHz would greatly
improve characterization of the spatial dependence of free-free and potential curvature in the synchrotron
SED.}

\end{enumerate}

We have estimated the synchrotron spectral index on degrees scales at a level currently not
reached by using typical spatial template methods alone,
or using parametric spectral fitting methods alone on a limited set
of maps (because the maps are insufficient for sampling
key frequency-dependent information).
Our results could be further improved by removal and inpainting
of point sources, and a more refined treatment of the 
free-free removal, possibly through incorporation of additional
frequencies in the spectral fitting.  New
low frequency sky surveys would provide key information by sampling the SED
at frequencies that currently lack data.  Future CMB
temperature observations at $\nu \lesssim 40$ GHz may
wish to consider frequency band choices that take advantage
of the frequency-pairing approach described here. 

We plan to make the ADR and $\beta_s$ maps available through the Legacy Archive for Microwave Background Data Analysis
(LAMBDA\footnote{\url{https://lambda.gsfc.nasa.gov}}) upon publication.

\vspace*{0.15in}
This research was supported in part by NASA grants 80NSSC21K0638, 80NSSC22K0408 and 80NSSC23K0475.
This research has made use of NASA's Astrophysics Data System Bibliographic Services. 
Some of the results in this paper have been derived using the \textsl{healpy} and \textsl{HEALPix} package.
We acknowledge the use of the 
Legacy Archive for Microwave Background Data Analysis (LAMBDA), part of the High Energy Astrophysics Science Archive Center (HEASARC). 
HEASARC/LAMBDA is a service of the Astrophysics Science Division at the NASA Goddard Space Flight Center.  
We also acknowledge use of the \textit{Planck} Legacy Archive. \textit{Planck} is an ESA science mission with instruments and contributions 
directly funded by ESA Member States, NASA, and Canada.
The Wisconsin H$\alpha$ Mapper and its H$\alpha$ Sky Survey have been funded primarily by the National Science Foundation. The facility was designed and built with the help of the University of Wisconsin Graduate School, Physical Sciences Lab, and Space Astronomy Lab. NOAO staff at Kitt Peak and Cerro Tololo provided on-site support for its remote operation.

\software{numpy \citep{harris/etal:2020}, scipy \citep{virtanen/etal:2020}, matplotlib \citep{hunter:2007}, astropy \citep{astropy2022}, HEALPix \citep{gorski/etal:2005}}

\appendix
\section{ADR monopole determination}
\label{app:monopoles}

Before a pixel-by-pixel parametric fit of spectral index may be performed using the Haslam 408 MHz, 23 GHz ADR and 30 GHz ADR maps, a Galactic emission monopole value must be determined for each frequency.  The main body of the text describes 
our choice of monopole for the Haslam map.  The monopoles for the ADR maps still need to be determined, however.

Since the Galactic emission of the ADR maps is dominated by synchrotron and free-free, we use a template-based method to determine the ADR map offsets:

\begin{equation}
    T_{ADR} = a T_{synch} + b T_{ff} + c
\end{equation}
where $T_{synch}$ is a synchrotron spatial template, and $T_{ff}$ is the free-free template.  The coefficent
$a$ can be translated into a mean synchrotron spectral index, and $b$ represents a conversion factor from
Rayleighs to temperature units.
This method assumes that the Galactic monopoles of the two templates are correct, and thus subtracting the offset $c$ from the ADR map will produce a consistent relative monopole normalization. 

We obtain an estimate of the free-free emission template $T_{ff}$ based on the all-sky velocity integrated WHAM
H$\alpha$ map.
The free-free template is derived from the H$\alpha$ map by applying a correction for extinction effects using the method described in
\cite{bennett/etal:2003, bennett/etal:2013}.  In the computation of the extinction optical depth 
$\tau$ ($=2.2 E(B-V)$) however,
we use the updated E(B-V) map of \cite{chiang:2023}
{\footnote{\url{https://lambda.gsfc.nasa.gov/product/foreground/fg_csfd_reddening_map_get.html}}}, 
which removes an extragalactic contribution from the originally used reddening map of \cite{sfd:1998}.
We also tried a version of the free-free template that included a scattering correction as described in  \cite{bennett/etal:2013}, but
the correction is not valid for $\tau \gtrsim 1$.  Differences between these two versions of the template did not significantly affect our results for higher latitudes with $\tau < 1$, and use of the de-extincted
template over the whole sky avoided introducing potential spatial discontinuities in maps due to the $\tau > 1$
limitation.

For the synchrotron template, we use the Haslam map after subtraction of an estimate of the free-free component at 408 MHz, based on an assumed power-law behavior for free-free.

The fit to determine the 23 and 30 GHz ADR monopoles is performed over a large fraction of the intermediate and high latitude sky.  We use a modified version of the analysis mask (see Section ~3.1) which already excludes bright point sources and bright diffuse Galactic emission within about $\pm 20^\circ$ of the plane. 
This portion of the fitting mask is chosen to avoid including regions of the sky affected by uncertainties in the foreground bias correction (see Section 3.1)
The mask is further modified with an additional conservative exclusion of the Galactic Haze region 
within $|\ell| < 60^\circ$,
since the Haze is not a prominent feature in the Haslam synchrotron template.

The chosen fitting mask effectively excludes regions of morphologically unique free-free emission, and so
does not permit an unconstrained evaluation of the free-free template coefficient over this region.  
We used a separate
set of less restrictive masks that thresholded on extinction optical depth ($0.3 <  \tau < 0.75$) to
obtain a range of acceptable values for the fitting coefficient $b$ at 23 GHz.  These values are mask
dependent, and fits returned a range for $b$ between
7.3 and 8.0 $\mu$K/R.  This range is consistent with quoted results in the literature.
\cite{harper/etal:2022} adopted a value of $195\pm5 \mu$K/R at 4.76 GHz, which, assuming a free-free spectral index of -2.12, extrapolates to 
a value of $7.24 \pm 0.19$ $\mu$K/R at 23 GHz. Earlier results from Planck \citep{planck/25:2015}
found $8.0 \pm 0.8 \mu$K/R.

We perform multiple fits of equation A1 to each of 23 and 30 GHz ADR maps by holding the values for $b$ fixed, consistent with values of [7.0, 7.2, 7.4, 7.6, 8.0] $\mu$K/R at 23 GHz, and solving for coefficients $a$ and $c$ as a function of $b$ and the chosen mask.
At 23 GHz, $-0.0525 \lesssim c_{23}  \lesssim -0.0515$ mK.  At 30 GHz, $c_{30} \sim -0.042$ mK.  Statistical errors are insignificant compared to systematic uncertainties caused by the choice of $b$
and template treatment. For both bands, $a$ returned values consistent with a synchrotron spectral index of
$-3.05$.  This is similar to typical values of $-3.04$ found by \citet{harper/etal:2022}.
For final values, we adopted $c_{23} = -0.052 $ mK and $c_{30} = -0.042$ mK in thermodynamic temperature, which most closely corresponded to the fixed
value of $b_{23} = 7.4 \mu$K/R.

\end{document}